\documentclass[9pt,epsf]{article}
\usepackage{CJKutf8}
\usepackage{amsmath}
\usepackage{booktabs, longtable}
\usepackage{caption2}
\usepackage{cite}
\usepackage{cmap}
\usepackage{enumerate}
\usepackage{fancybox}
\usepackage{flafter}
\usepackage{indentfirst}
\usepackage{latexsym}
\usepackage{listings}
\usepackage{pxfonts}
\usepackage{textcomp}
\usepackage{times}
\usepackage[T1]{fontenc}
\usepackage[dvips]{graphicx}
\usepackage{titletoc}
\usepackage{authblk}
\usepackage{multirow}
\usepackage{subfigure}
\usepackage{hyperref}
\usepackage{color}
\usepackage{graphicx}  
\usepackage{float}     
\usepackage{yhmath}
\usepackage{url}
\usepackage{appendix}
\usepackage{xcolor}
\usepackage{color}
\usepackage{flushend, cuted}
\usepackage{mathrsfs}

\definecolor{lightgray}{rgb}{.7,.7,.7}

\definecolor{red}{rgb}{1,0,0}

\definecolor{blue}{rgb}{0,0,1}

\begin{document}

\begin{center}
{\Large Phase-Transition Theory of Kerr Black Holes in Electromagnetic Field }
\vskip 1.0cm

\centerline {~Yi Liao$^{a}$
\footnote{liaoy@mail.sustc.edu.cn},
Xiao-Bo ~Gong $^{b}$
\footnote{gxbo@ynao.ac.cn},
and Jian-Sheng ~Wu$^{a}$
\footnote{\emph{Corresponding author}: wu.js@sustc.edu.cn}}
{\small 
$^a$ Department of Physics, South University of Science and Technology of China, Shenzhen, 518055, China\\
$^b$ Yunnan Observatory, Chinese Academy of Sciences, Kunming, 650011, China}
\end{center}
\vskip 1.5cm
\begin{abstract}
For a Kerr black hole (KBH) with spin $J$ and mass $M$ in a steady electromagnetic field, a special Wald vacuum solution (WVS) has been found in the case of no-source uniform field.  For WVS, the Meissner effect (ME) occurs only in the the extreme KBH where $M^2/J=1$, in this case,  the  magnetic field is totally excluded from the event horizon (EH) of KBH.  However, WVS does not consider the Hawking radiation (HR) but treats KBH as an absolutely black body.  If HR is added , researchers believe that the condition is not so restricted and it is possible for ME to occur in less extreme case.   How less is the "less extreme case"? This paper tries to answer this question. Since the Hawking temperature $T_H$ of KBH   defined by HR  is proportional to the surface gravity $\kappa$ at the EH,  this question is actually  about the so-called existence/non-existence of ME (ME/NME)  or superconducting phase transition. In this paper, we study the connection between the superconductivity of KBH-EH and the existence of Weyl Fermion (WF). Using thermodynamic formulas and the KBH state equation, we prove that the inherent-parameter condition for ME to occur is  $M^2/J\leq \epsilon_c=1.5$ in force-free fields whether it be in the simple axisymmetric vacuum zero source case or in the non-zero source case which can be described by the nonlinear Grad-Shafranov (G-S) equation. We suggest that this is a second-order phase transition and we calculate the critical exponents $\delta=1$ and $\eta=1/2$ for the specific heat diverging at constant $J$, and the critical point $(M_c, \Omega_c)$, which equals $(1.22\sqrt{ J}, 0.16/\sqrt{ J})$ where $\Omega$ is the angular velocity of KBH. Furthermore we draw the phase diagrams in both $(M, J)$ and $(M,\Omega)$ coordinates.
\end{abstract}

\newpage
{\section{Introduction}}
Black hole thermodynamics has been an intriguing subject of discussions for decades. The analogy between space-time with Black hole horizon and thermodynamics have been extensively investigated \cite{Hawking1973}. The four laws of black hole thermodynamics were discovered by Carter, Hawking, and Bardeen \cite{Hawking1973}. These laws are physical properties that black holes are believed to satisfy. A black hole behaves as a blackbody with the Hawking temperature $T_H=\frac{\hbar \kappa}{2\pi k_Bc}$, where $\kappa$, $c$, $k_B$, and $\hbar$ denote the surface gravity of the event horizon (EH) , light velocity, Boltzmann constant,  and Planck constant respectively \cite{Hawking1974,Hawking1975}. Since thermodynamic quantities correspond to microscopic structures, some researchers have made attempt to explain Black hole properties through  exploring  phase transition of the small-large charged Schwarzschild black hole(SBH/LBH) \cite{Hawking,Chamblin1,Chamblin2,Wei}. This kind of phase transitions are similar to the liquid-gas transitions which are first-order \cite{Liu}. Compared with Schwarzschild black hole, there might exist more complex phenomena in the Kerr black hole (KBH) since it has additional spin $J$.

Blandford $\&$ Znajek (1977) gave the nonlinear Grad-Shafranov(G-S) equation for a force-free magnetosphere (FFM) of the curved Kerr space-time, which can describe the energy extraction process\cite{Blandford}.  In this progress, KBH can be heated or cooled to reach a phase transition. One recalls Boltzmann’s insight: “If you can heat it, it has microscopic structure.” The rotational energy of the KBH can be converted into the thermal and kinetic energy of the surrounding plasma. The electron would emit many photons, which in turn can produce a plentiful supply of the electron-positron pairs. When a KBH is immersed in a steady electromagnetic field, for the Wald vacuum solution (WVS)\cite {Wald}, King et al. (1975) found that all magnetic fields are expelled out of the event horizon of the extremal KBH \cite{King}. The WVS is time-independent and it describes a KBH being immersed in a uniform magnetic field aligned with the black hole spin axis. Bi{\v c}{\'a}k \& Dvo{\v r}{\'a}k(1976) and Bi{\v c}{\'a}k \& Janis(1985) generalized this result \cite{Bicak, Bi}. They showed that if the black hole is extremal ( namely $M/a=1$, where $M$ denotes the mass of KBH and $a=J/M$), non-monopole component of magnetic flux could not penetrate the event horizon for all steady axisymmetric vacuum solutions. The interesting phenomenon is called Meissner effect (ME), which is the expulsion of magnetic field lines out of the event horizon and the quenching of jet power for the KBH \cite{Bicak, Penna}.  However, WVS does not consider the Hawking radiation(HR) but treats KBH as an absolute blackbody\cite{Hawking1974,Wald}.  If the HR is considered , researchers believe that the condition is not so restricted and it is possible for ME to occur in less extreme case\cite{Penna}. Some observations about astrophysical black holes with high spins support this argument\cite{McClintock2006,Gou,McClintock2014}. The spin parameters of the near-extreme Kerr black holes in Cyg X-1 and GRS 1915 + 105 have been measured to be $M^2/J$ greater than 1. Jet from these black holes could be quenched by the ME.

Recently, for  zero or non-zero source, the ME in the steady axisymmetric FFM attracts lots of attentions  where HRs have been considered. Many physicists want to know whether the Blandford-Znajek (BZ) process would be quenched if the Meissner effect exists in the real astrophysical environment. However, this effect was never seen in the previous general relativistic magnetohydrodynamic simulations. In our previous work \cite{Gong}, we discuss about the external-field condition for ME to occur which is $H_{\phi}=0,\omega = \Omega_0 \equiv \frac{a}{a^2+r_{+}^2}$ on the external EH. Here $\Omega_0$ is the angular velocity of KBH-EH, ${H_{\phi}}$ corresponds to  twice the poloidal electric current, and $\omega=-A_{0,r}/A_{\phi,r}=-A_{0,\theta}/A_{\phi,\theta}$ which satisfies $\vec{E}=-\vec{\omega}\times \vec{B}$, with $A_i$ and $r_+$ being the $i$th component of vector potential $\vec{A}$ and the radius of external EH \cite{Blandford, Znajek},  we denote $\partial X/\partial y$ as $X_{,y}$. On the other hand, Znajek (1978) suggested that the effective electric resistance of KBH-EH surface is a non-zero constant indicated by G-S equation if the BZ mechanism works \cite{Blandford,Znajek1978}. In Ref.\cite{Gong}, we prove that BZ mechanism only works when the angular velocity  $\omega$ of the field is strictly slower than the angular velocity $\Omega_0$ of KBH. If the BZ mechanism does not work, the resistance may vanish. It could be regarded as a superconducting phase transition since the electric resistance suddenly vanishes. 

In the paper, we study the connection between the superconductivity of KBH-EH and the existence of Weyl Fermion (WF). We also give the thermodynamic KBH state equation and the inherent-parameter condition for the existence/non-existence of ME (ME/NME) phase transition where $M^2/J=\epsilon_c=1.5$. The divergence of the specific heat at constant $J$ tells that this is a second-order phase transition. Therefore we also calculate the critical exponents and find the critical points, further we draw the phase diagrams in $(M,J)$ and  $(M,\Omega)$ coordinates.  Moreover, we provide a possible explanation that the parameters of ME phase exactly correspond to positive specific heat condition. It is a necessary condition to reach holographic superconductivity\cite{Horowitz}.

For simplicity, the constant $\hbar$,  $k_B$, $c$, and the gravitational constant $G$ equals unit $1$.
{\section{Meissner effect }\label{se:Meissiner}}
The Kerr metric in Boyer-Lindquist coordinates is
\begin{eqnarray}
 ds^{2}&=&g_{\mu\nu}dx^{\mu}dx^{\nu} \nonumber
 \\ &=&-(1-\frac{2Mr}{\Sigma})dt^{2}-\frac{4Mar~{\rm \sin^{2}}\theta}{\Sigma}dtd\phi+\frac{\Sigma}{\Delta}dr^{2}
 +\Sigma d\theta^{2}+\frac{A~{\rm \sin^{2}}\theta}{\Sigma}d\phi^{2}.
\label{eq:metric}
\end{eqnarray}
Here $x^{\nu}=(t,r,\theta,\phi), ~~ \nu=(0,1,2,3) \nonumber, \Delta=r^{2}-2Mr+a^{2}, \Sigma=r^{2}+a^{2}\cos^{2}\theta$ and
 $A=  (r^{2}+a^{2})\Sigma+2Mra^{2}\sin^{2}\theta=(r^{2}+a^{2})^{2}-\Delta a^{2}\sin^{2}\theta$.

The constraint differential equation for the FFM around KBHs is given by Menon $\&$ Dermer (2005)\cite{Menon} as
\begin{equation}
\begin{aligned}
 \frac{\sqrt{\gamma}}{2\alpha\gamma_{\phi\phi}}\frac{dH_{\phi}^{2}}{dA_{\phi}}=&\omega\partial_{r}[\frac{\gamma_{\theta\theta}}{\alpha
 \sqrt{\gamma}}(\gamma_{\phi\phi}\omega+\beta_{\phi})A_{\phi,r}]+\omega\partial_{\theta}[\frac{\gamma_{rr}}{\alpha
 \sqrt{\gamma}}(\gamma_{\phi\phi}\omega+\beta_{\phi})A_{\phi,\theta}]     \\
 &+\partial_{r}[\frac{\gamma_{\theta\theta}}{\alpha\sqrt{\gamma}}(\beta^{2}-\alpha^{2}+\beta_{\phi}\omega)A_{\phi,r}]+
 \partial_{\theta}[\frac{\gamma_{rr}}{\alpha\sqrt{\gamma}}(\beta^{2}-\alpha^{2}+\beta_{\phi}\omega)A_{\phi,\theta}],
 \end{aligned}
\label{eq:constraint}
\end{equation}
where $\alpha=\sqrt{{\frac{\Delta\Sigma}{A}}}, \beta_{\phi}=-\frac{2Mra}{\Sigma}{\rm \sin^{2}}\theta,
\sqrt{\gamma}=\sqrt{\frac{A\Sigma}{\Delta}}{\rm \sin}\theta,
\gamma_{\phi\phi}=\frac{A{\rm \sin^{2}}\theta}{\Sigma}, \gamma_{\theta\theta}=\Sigma, \gamma_{rr}=\frac{\Sigma}{\Delta},
\beta^{2}-\alpha^{2}=\frac{2Mr}{\Sigma}-1$.
 The angular velocity $\omega(A_{\phi})$ is a function of $A_{\phi}$.
 Considering the Euler-Lagrange equation, which is $\partial_{\nu}\frac{\partial\mathscr{L}}{\partial A_{\phi,\nu}}=\frac{\partial \mathscr{L}}{\partial A_{\phi}}$,
 the Lagrangian $\mathscr{L}$ for Eq.(\ref{eq:constraint}) can be expressed as\cite{Gong}
\begin{small}
\begin{equation}
\begin{aligned}
 \mathscr{L}
&=\frac{1}{\sin \theta}(g_{33}\omega^{2}+2g_{03}\omega+g_{00})(A_{\phi,r}^{2}+\frac{1}{\Delta}A_{\phi,\theta}^{2})
  &+\frac{1}{\sin \theta}g_{11}H_{\phi}^{2}.
   \end{aligned}
\label{eq:lagrangian}
\end{equation}
\end{small}
  One can get the external and inner light surfaces for the given $\omega$ by solving
\begin{equation}
  L\equiv g_{33}\omega^{2}+2g_{03}\omega+g_{00}=0.
\end{equation}
 Namely $(\omega-\Omega')\varpi=\pm\alpha$, where $\Omega'=\frac{2aMr}{A}$,
  $\varpi=\sqrt{\frac{A}{\Sigma}}\sin\theta$. Let $\mu={\rm -\cos\theta}$, and we assume that $L(A_{\phi},r,\mu)$ is a function of $A_{\phi},~r,~\mu$. Then Eq.(\ref{eq:constraint}) becomes G-S equation,which reads
\begin{small}
\begin{equation}
\begin{aligned}
L(\frac{A_{\phi,rr}}{1-\mu^{2}}+\frac{A_{\phi,\mu\mu}}{\Delta})+
\frac{L{_{,r}}A_{\phi,r}}{1-\mu^{2}}+\frac{L{_{,\mu}}A_{\phi,\mu}}{\Delta}+
\frac{1}{2}L{_{,A_{\phi}}}(\frac{A_{\phi,r}^{2}}{1-\mu^{2}}+\frac{A_{\phi,\mu}^{2}}{\Delta})-
\frac{g_{11}H_{\phi}}{1-\mu^{2}}\frac{dH_{\phi}}{d A_{\phi}}=0.
\end{aligned}
\label{eq:gsequation}
\end{equation}
\end{small}
Here, $i=r$ or $\mu$, $L{_{,i}}=g_{33,i}\omega^{2}+2g_{03,i}\omega+g_{00,i}$, and $~L{_{,A_{\phi}}}=2(g_{33}\omega\frac{d\omega}{d A_{\phi}}+g_{03}\frac{d\omega}{d A_{\phi}})$.

The Lorentz invariant ${\frac{1}{2}\rm \mathscr{F}^{2}}=\frac{1}{2}F_{\mu\nu}F^{\mu\nu}$ to the Carter observers is
$B^{2}-E^{2}$, where the Carter field components given by Znajek (1977)\cite{Znajek} are
\begin{equation}
\begin{aligned}
&E_{r}=[a-\omega(r^{2}+a^{2})]A_{\phi,r}/\Sigma,
\\&B_{r}=[(1-a\omega{\rm \sin^{2}}\theta)]A_{\phi,\theta}/(\Sigma{\rm \sin}\theta), \\
&E_{\theta}=[a-\omega(r^{2}+a^{2})]A_{\phi,\theta}/(\Sigma\sqrt{\Delta}),
\\&B_{\theta}=-[\sqrt{\Delta}(1-a\omega{\rm \sin^{2}}\theta)]A_{\phi,r}/(\Sigma{\rm \sin}\theta), \\
&E_{\phi}=0,
\\&B_{\phi}=H_{\phi}/(\sqrt{\Delta}{\rm \sin}\theta).
\end{aligned}
\label{eq:fieldcomponts}
\end{equation}
Then we can get current density $j_{\mu}\equiv g_{\mu\nu}j^{\nu}=F_{\mu\nu;\nu}$ ($;$ denotes covariant derivative).
Specially, for the  vacuum case which means $j_{\mu}\equiv 0$, the electromagnetic field vanishes at the EH if ME occurs. There is no charged particle produced by Hawking radiation.

For G-S equation, one can get the electric current density in the $(t,r,\theta,\phi)$ coordinates from  Eq.(\ref{eq:fieldcomponts}), which reads,
\begin{equation}
\begin{aligned}
&j^{0}=-\frac{1}{\Sigma}\left\{[\frac{(g_{33}\omega+g_{03})A_{\phi,r}}{1-\mu^{2}}]_{,r}+
[\frac{(g_{33}\omega+g_{03})A_{\phi,\mu}}{\Delta}]_{,\mu}\right\},
\\&j^{r}=g^{-\frac{1}{2}}\frac{dH_{\phi}}{d A_{\phi}}A_{\phi,\theta},
\\&j^{\theta}=-g^{-\frac{1}{2}}\frac{dH_{\phi}}{d A_{\phi}}A_{\phi,r},
\\&j^{\phi}=\omega j^{0}+g^{-\frac{1}{2}}\frac{H_{\phi}\Sigma}{\Delta {\rm \sin}\theta}.
\end{aligned}
\label{eq:fieldcompontsdesity}
\end{equation}
 Here, $g$ is the determinant of  the metric tensor ${g_{\mu\nu}}$. When $\omega=\Omega_0$, the boundary condition $H_{\phi}=\frac{{\rm \sin}\theta[\omega(r_{+}^{2}+a^{2})-a]}{r_{+}^{2}+a^{2}\cos^{2}\theta}A_{\phi,\theta}$ at the EH leads to  $H_{\phi}=0$. From Eq.(\ref{eq:fieldcomponts}), one can know that the electromagnetic field is zero at the EH.
 From Eq.(\ref{eq:fieldcompontsdesity}), one can know that $j^{r}=0$ at the EH when the ME occurs.  If $\vec{j}\neq \vec{0}$, it means the resistance is zero. The charged particles produced by Hawking radiation cannot drop into the KBH.
  We think the charged particle with light velocity can stay on the EH. If the current density $\vec{j}$ is a null-vector (namely $j_\mu j^\mu=0$)
and $\vec{j}\neq \vec{0}$, it means that the current carrier is a charged particle with light velocity, whose mass should be nought. When the interaction between them are very weak, the resistance should be vanishing\cite{Punsly}. The most possible candidate serving as the carrier is the mysterious Weyl Fermion (WF) which is massless and permitted to carry charges. If the carrier is an ordinary charged particle with mass, the null-vector $\vec{j}$ must satisfy $j_{\mu}\equiv 0$. In other words, the electric current at the external EH vanishes. Thus, in theory we can test the existence of WF by measuring if there is a nonvanshing current on the EH surface when $\omega=\Omega_0$ . Moreover, because the resistance of particles with mass hardly vanishes \cite{Punsly}, if the superconductivity of KBH-EH exists, so does WF. It should be pointed out that the above arguments are not explicitly based on the quantum effect. However since the thermal behaviors of the black hole are essentially quantum mechanical, so are its electric behaviors. The quantum fluctuations might change this result, the thorough answer to this problem may require a better self-consistent theory of quantum gravity \cite{Natusuume}.

The final result of the Lorentz invariant is\cite{Gong}
\begin{small}
\begin{equation}
\begin{aligned}
{\frac{1}{2} \rm \mathscr{F}^{2}}&=\frac{1}{\sqrt{-g}}(\frac{2}{\sin \theta}\frac{\Sigma}{\Delta}H_{\phi}^{2}-\mathscr{L}).
\end{aligned}
\label{eq:f2}
\end{equation}
\end{small}
So $\mathscr{L}=-\sqrt{-g}[(B_{r}^{2}+B_{\theta}^{2}-B_{\phi}^{2})-(E_{r}^{2}+E_{\theta}^{2}+E_{\phi}^{2})]$. This relation can
be found in MacDonald $\&$ Thorne (1982)\cite{MacDonald}. Komissarov (2004) analyzed this poloidal electric field in detail at this region\cite{Komissarov}.

In Ref.\cite{Gong}, for G-S equation, we prove that the ME expels magnetic fields out of the external EH and it also expels magnetic fields out of the inner light surface. The magnetic fields outside the inner light surface could not go to the region between the external EH and the inner light surface. To discuss the comprehensive phase transition, taking into account the condition where $BZ$ mechanism becomes invalid, we could conclude as follows: for any angular velocity $\Omega$ determined by an equilibrium state equation, a necessary external-field condition for ME to occur is $\omega=\Omega$ at the external EH.

{\section{Phase transition }\label{se:phase}}
For KBH, the external and inner EH exists at $ r_{\pm}=M\pm \sqrt{M^{2}-a^{2}}$. And we notice that the surface gravity $\kappa $ of KBH reads
 $\kappa=\frac{r_+-r_-}{2(a^2+r_{+}^2)}=\frac{\sqrt{M^2-a^2}}{a^2+r_{+}^2}$. So we have the equation of the mass function $M(\kappa, J)$ satisfying
\begin{equation}
  \kappa =\frac{\sqrt{M^4-J^2}}{2M(M^2+\sqrt{M^4-J^2})}.
  \label{eq:temperature}
\end{equation}
At constant $J$, when $M=J$, $\kappa=0$; when $M=+\infty$, $\kappa=0$. Thus, for the same surface gravity $\kappa$, there exist different mass $M$.
Meanwhile, we have the first law of KBH thermodynamics in the electromagnetic field, which is $dM=dW+\Omega dJ+ \kappa dA_{+}$, with $A_{+}$  the external surface area of EH. Here $dW$ is the rotation energy extracted by the field. Punsly(2008) has showed that axisymmetic vacuum fields could not extract energy from a black hole\cite{Punsly,Punsly1989}, namely $dW=0$. For G-S equation, $dW=\dot{W}dt$, the effective power $\dot{W}$ of electromagnetic field affecting KBH satisfies\cite {Blandford}
$\dot{W}\propto \omega(\Omega-\omega)$.
When  $\omega=\Omega$, $\dot{W}=0$. Therefore we have the first law of KBH thermodynamics that reads
\begin{equation}
  dM=\Omega dJ+ \kappa  dA_{+}.
\end{equation}
Here, $M$, $-\Omega$, $J$, and $A_{+}$ is analogous to the internal energy, pressure, volume and entropy of gas\cite{Ho}.
The field does not affect the state of KBH, so we have $ \Omega= \Omega_0=\frac{a}{a^2+r_{+}^2}.$
The thermodynamic state equation $\Omega (\kappa, J)=0$  reads
\begin{equation}
  (2M{\Omega})^2+(4M{\kappa})=1;\kappa =\frac{\sqrt{M^4-J^2}}{2M(M^2+\sqrt{M^4-J^2})}.
\end{equation}

\begin{figure}
 \centering
 \includegraphics[angle=0,scale=0.55,bbllx=20pt,bblly=200pt,bburx=800pt,bbury=100pt]{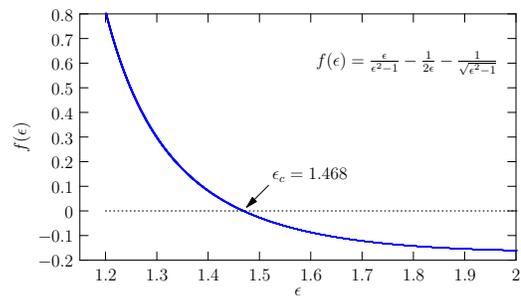}
    \caption{The curve $f(\epsilon)$,with $\epsilon=M^2/J$, which determines the critical value $\epsilon_c\approx 1.5$ where specific heat diverges.
}
   \label{fig:zero}
\end{figure}

Let $\epsilon \equiv \frac{M^2}{J},M=\sqrt{J\epsilon}$, $\epsilon\geq 1$,we have the variable $\epsilon$ satisfying
\begin{equation}
  \ln \kappa +\frac{1}{2}\ln J+\ln2 =\frac{1}{2}[\ln(\epsilon^2-1)-\ln \epsilon)]-\operatorname{arccosh}\epsilon.
\label{eq:derviative}
\end{equation}
The specific heat at constant spin $J$ is
\begin{equation}
  C_{J}\equiv (\frac{\partial M}{\partial \kappa})_{J}=\frac{J}{2M}(\frac{\partial \epsilon}{ \partial \kappa })_{J}.
  \label{eq:heat}
\end{equation}
Differentiating Eq.(\ref{eq:derviative}) with respect to $\kappa$, we have
$
  (\frac{\partial \epsilon}{ \partial \kappa })_{J}^{-1}=\kappa[(\frac{\epsilon }{\epsilon^2-1})-(\frac{1}{2\epsilon}+\frac{1}{\sqrt{\epsilon ^2-1}})].
$
Apart from the trivial diverging points ($\kappa=0$ and $\epsilon=+\infty $), the singular points of $C_{J}$ correspond to the roots of the function $f(\epsilon)$ which can be written as
\begin{equation}
 f(\epsilon)\equiv (\frac{\epsilon }{\epsilon^2-1})-(\frac{1}{2\epsilon}+\frac{1}{\sqrt{\epsilon ^2-1}})=0.
\end{equation}
The reciprocal of specific heat satisfies
\begin{equation}
  C_{J}^{-1}=\frac{2M\kappa}{J}f(\frac{M^2}{J}).
  \label{eq:reciprocal}
\end{equation}
As shown in Fig.\ref{fig:zero}, the zero point is $\epsilon_c\approx 1.5$ by numerical calculation.

The critical point $(M_c,\Omega_c)$ at constant $J$ satisfies
 \begin{equation}
   M_c=\sqrt{\epsilon_c J}=1.22\sqrt{J}, \Omega_c=\frac{1}{2}\sqrt{\frac{\epsilon_c-\sqrt{\epsilon_c^2-1}}{\epsilon_c(\epsilon+\sqrt{\epsilon_c^2-1})}}\frac{1}{\sqrt{J}}=\frac{0.16}{\sqrt{J}}.\textit{}
 \end{equation}
  The critical  surface gravity $\kappa_c$ reads
 \begin{equation}
 \kappa_c =\frac{\sqrt{\epsilon_c^2-1}}{2\sqrt{\epsilon_c}(\epsilon_c+\sqrt{\epsilon_c^2-1})}\frac{1}{\sqrt{J}}=\frac{0.174}{\sqrt{J}}=\frac{0.213}{M_c}.
 \end{equation}

 The critical exponent $\delta$ and $\eta $ is defined by
 \begin{equation}
  |C_{J}^{-1}(\tilde{M})|\sim |\tilde{M}|^\delta , |C_{J}^{-1}(\tilde{\kappa})|\sim |\tilde{\kappa}|^\eta.
 \end{equation}
  Here, the reduced surface gravity $\tilde{\kappa}$ and mass $\tilde{M}$ satisfies $\kappa\equiv \kappa_c(1+\tilde{\kappa})$, $M \equiv M_c(1+\tilde{M})$. Then we let $J=1$, so $\epsilon=M^2$, $M_c=1.224$ and $\kappa_c=0.174$.
 From Eq.(\ref{eq:reciprocal}), the reciprocal of specific heat  $C_{1}$ satisfies
 \begin{equation}
    C_{1}^{-1}(\tilde{M})=2.45\kappa(1+\tilde{M})f(1.5(1+\tilde{M})^2)\equiv C_{1}^{-1}(\tilde{\kappa})=0.35(1+\tilde{\kappa})Mf(M^2).
   \end{equation}
   When $\tilde{M}<<1$, near critical point $M_c$, we can re-write it as $C_{1}^{-1}(\tilde{M}) \approx 0.7 [1.2(1-5\tilde{M})-0.3(1-2\tilde{M})-0.9(1-3.6\tilde{M})] \approx -1.5\tilde{M}\sim \tilde{M}$. Meanwhile, we have $\tilde {\kappa}\approx \frac{1+3.6\tilde{M}-1.1\tilde{M}^2}{1+3.6\tilde{M}+2.7\tilde{M}^2}-1\approx -3.8\tilde{M}^2\sim \tilde{M}^2$. So, $ |C_{1}^{-1}(\tilde{\kappa})|\sim |\tilde{\kappa}|^{1/2}$ near point $\kappa_c$.
 Therefore, the exponent $\delta=1$ and $\eta=\frac{1}{2}$.

\begin{figure}
 \centering
 \includegraphics[angle=0,scale=0.35,bbllx=20pt,bblly=200pt,bburx=450pt,bbury=100pt]{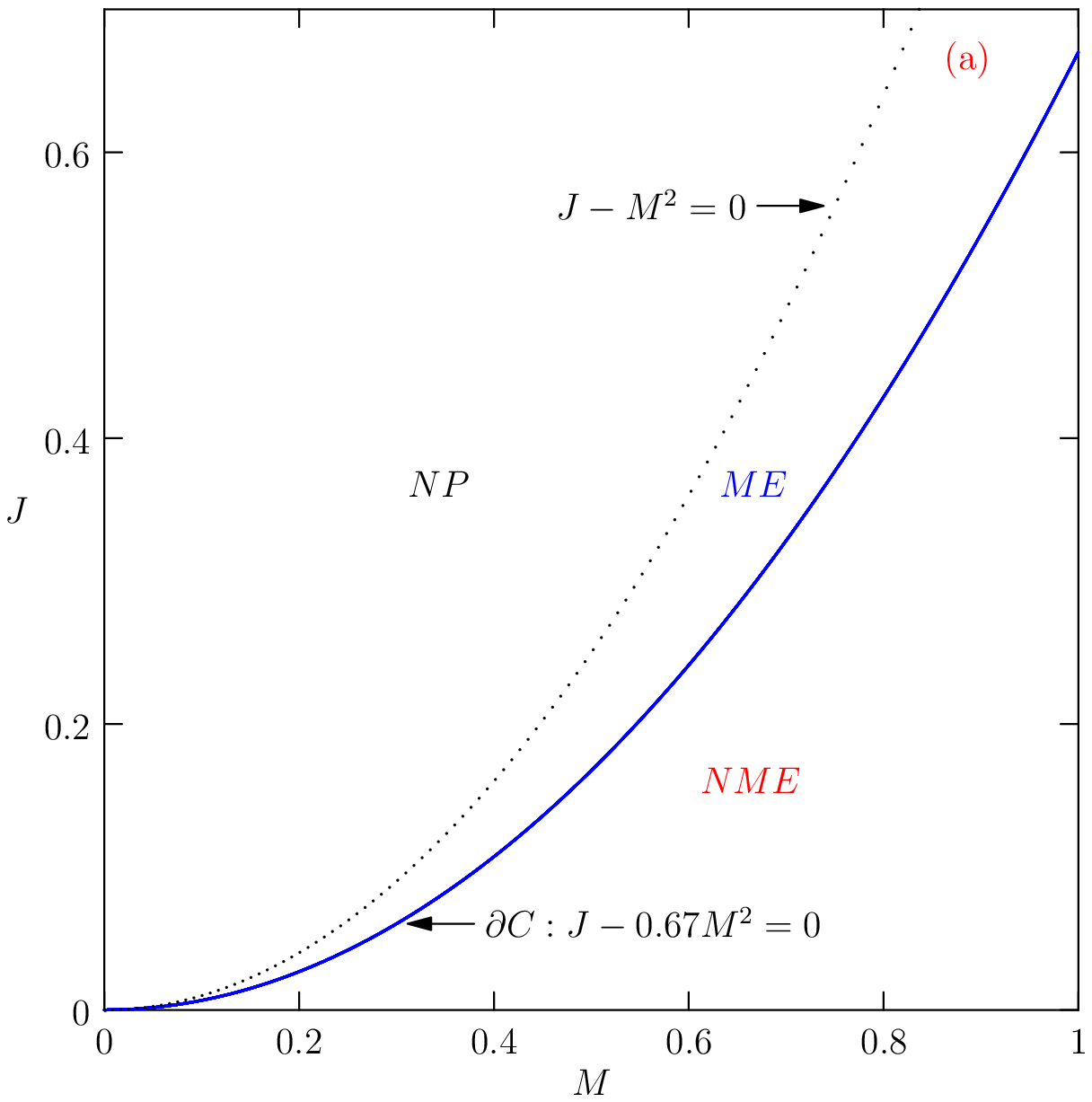}
  \includegraphics[angle=0,scale=0.35,bbllx=20pt,bblly=200pt,bburx=450pt,bbury=100pt]{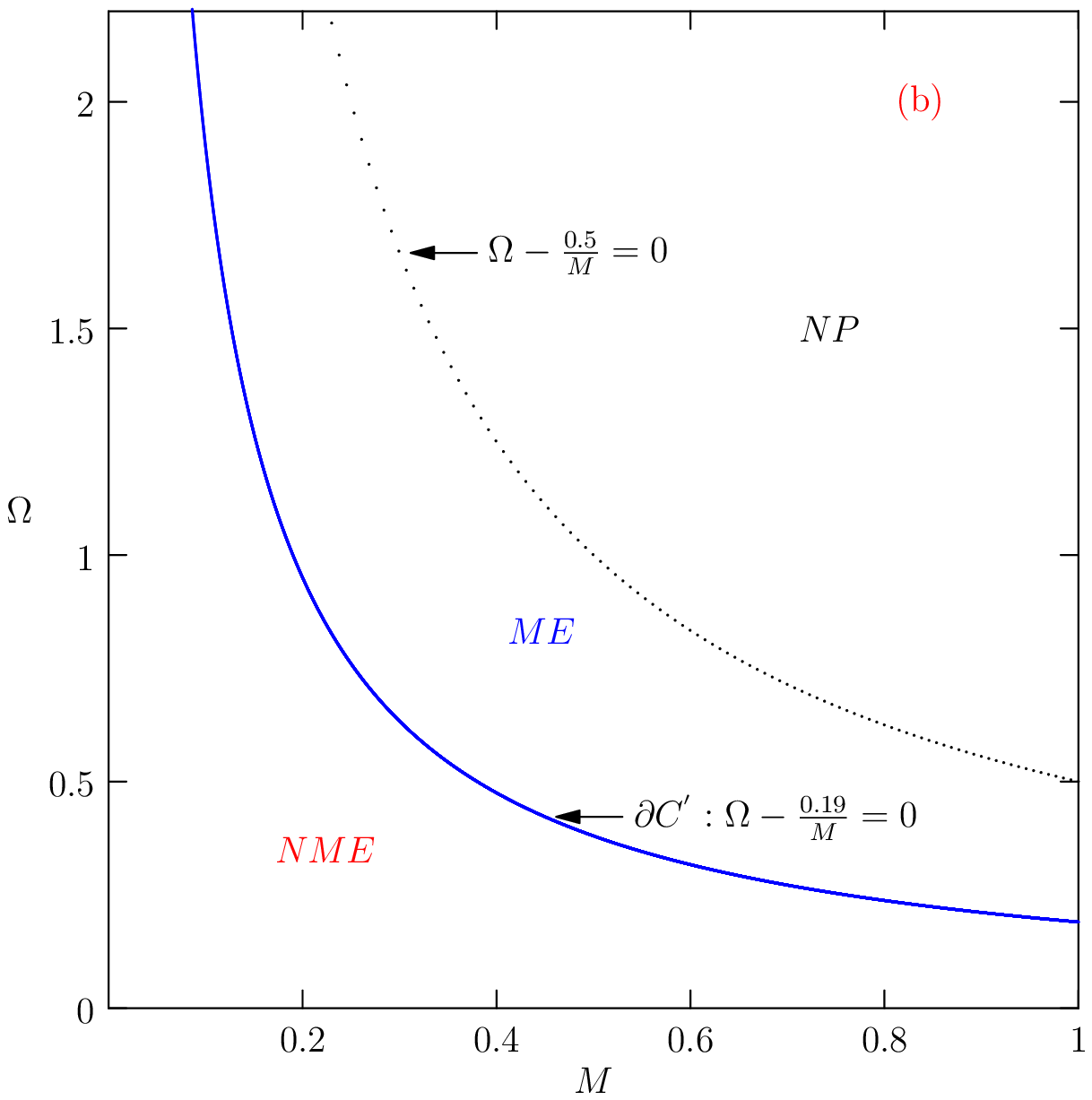}
    \caption{The left panel:(a). phase diagram in $(M, J)$ coordinates; The right panel: (b). phase diagram in $(M,\Omega)$ coordinates. NP represents the non-physical zone.
     }
     \label{fig:phase}
\end{figure}

  As shown in Fig.\ref{fig:phase}, in ME/NME phase diagram, the critical curve $\partial C$ in $(M,J)$ coordinates and the corresponding curve $\partial C'$ in $(M,\Omega)$ coordinates reads
  \begin{equation}
    \partial C:J-0.67M^2=0.\& ~\partial C': \Omega-\frac{0.19}{M}=0.
  \end{equation}

 One can notice that the parameters of ME phase exactly correspond to the positive specific heat condition. It is a necessary condition to reach holographic superconductivity\cite{Horowitz}. In the superconductor, we need condensation of matter field yielded by Hawking radiation. A nonzero condensate corresponds to a static nonzero field outside the black hole. This viewpoint is helpful to understand the physical cause of the phase transition. We will focus on the proof of superconductivity in our future works.

{\section{Summary and discussion }\label{se:sum}}

We give the thermodynamic state equation of KBH, calculate the condition of the ME/NME phase transition and draw the phase diagrams. We also get the critical exponents $\delta=1$ and $\eta=\frac{1}{2}$. It needs to be pointed out that the ME/NME phase transition is only a terminology, like SBH/LBH phase transition. Lacking the knowledge about microscopic structure of KBH, it is hard to tell whether ME is the  essential criterion for the phase transition or not. Comparing with Bardeen-Cooper-Schrieffer(BCS) superconductvity, there may exist a more natural phenomenon where the electric resistance vanishes. However the phase transition suggests that there exists a kind of massless and charged particle, possibly WF, which waits to be confirmed in the future. If ME and zero-resistance occurs simultaneously when BZ mechanism does not work, one can study if the critical magnetic field exists for the linear vacuum case or more complex cases.  Besides, since ME/NME phase transition is second order, one could find the order parameter and construct the corresponding Landau-Ginzburg equation. On the other hand, Anti-de Sitter(AdS) KBH with negative cosmological constant $\Lambda $ or Kerr-Newman black hole with charge $Q$ may exhibit richer phase transitions. It is very difficult to find the condition for ME to occur in these cases. The present calculation for phase transition of KBH serves as the first step to explore the above mentioned problems, which will be studied in our future works. We are looking forward to understanding these phenomena in the future researches.
{\section*{Acknowledgments}}
This work was supported in part by startup funding of the South University of Science and Technology of China, the Shenzhen Peacock Plan and Shenzhen Fundamental Research Foundation (Grant No. JCYJ20150630145302225).
\newpage 
Fig.\ref{fig:zero}
\newpage 
Fig.\ref{fig:phase}

\end{document}